\documentclass[aps,prl,twocolumn,preprintnumbers,linenumber,amsmath,amssymb,floatfix]{revtex4-1}
\usepackage{graphicx}
\usepackage{tabularx}
\usepackage{subfigure}
\usepackage{epsfig}
\usepackage{dcolumn}
\usepackage{bm}
\usepackage{euscript}
\usepackage{ulem}
\usepackage{color}
\usepackage{epsfig,psfrag,subfigure}
\usepackage{amsfonts}
\usepackage{exscale}
\usepackage{amsbsy}
\usepackage{ulem}
\usepackage{lineno}
\usepackage{blindtext}

\def\avg#1{\left\langle#1\right\rangle}

\def\ket#1{\left|#1\right\rangle}

\def\be{\begin{equation}}       \def\ee{\end{equation}}
\def\bea{\begin{eqnarray}}      \def\eea{\end{eqnarray}}
\def\ba{\begin{array}}
\def\ea{\end{array}}
\def\bnum{\begin{enumerate} }
\def\enum{\end{enumerate}}

\def\nn{\nonumber}

\def\=>{\Rightarrow}
\def\>{\rightarrow}

\def\eye2{Fathbb{I}}

\def\Eq#1{Eq.~(\ref{#1})}
\def\Fig#1{Fig.~\ref{#1}}

\renewcommand{\>}{\rangle}

\begin{document}
\title{Fermion-induced quantum critical points}

\author{Zi-Xiang Li$^{1}$, Yi-Fan Jiang$^{1}$, Shao-Kai Jian$^{1}$, Hong Yao$^{1,2,3,\ast}$}
\affiliation{
$^1$Institute for Advanced Study, Tsinghua University, Beijing 100084, China\\
$^2$State Key Laboratory of Low Dimensional Quantum Physics, Tsinghua University, Beijing 100084, China\\
$^2$Collaborative Innovation Center of Quantum Matter, Beijing 100084, China
}

\begin{abstract}
A unified theory of quantum critical points beyond the conventional Landau-Ginzburg-Wilson paradigm remains unknown. According to Landau cubic criterion, phase transitions should be first-order when cubic terms of order parameters are allowed by symmetry in the Landau-Ginzburg free energy. Here, from renormalization group (RG) analysis we show that second-order quantum phase transitions can occur at such putatively first-order transitions in interacting two-dimensional Dirac semimetals. As such type of Landau-forbidden quantum critical points are induced by gapless fermions, we call them fermion-induced quantum critical points (FIQCP). We further introduce a microscopic model of SU($N$) fermions on the honeycomb lattice featuring a transition between Dirac semimetals and Kekule valence bond solids. Remarkably, our large-scale sign-problem-free Majorana quantum Monte Carlo simulations show convincing evidences of a FIQCP for $N=2,3,4,5,6$, consistent with the RG analysis.  We finally discuss possible experimental realizations of the FIQCP in graphene and graphene-like materials.
\end{abstract}

\date{\today}
\maketitle

Fathoming the behavior of quantum matters near quantum phase transitions in strongly correlated many-body systems is among central and challenging issues in modern condensed matter physics \cite{Subirbook}. Due to Landau and Ginzburg \cite{Landau-99}, a prevalent understanding of phase transitions is provided by order parameters whose nonzero expectation value can characterize phases with lower symmetries. Sufficiently close to the transition point, order parameter fluctuations at large distances and long times dominate the physics near such phase transitions and are described by a continuum field theory of order parameters. Combined with Wilson's renormalization group (RG) theory \cite{Wilson-74}, this sophisticated Landau-Ginzburg-Wilson (LGW) paradigm for phase transitions has made huge successes in understanding second-order phase transitions in correlated many-body systems including superconductors, density-wave compounds, and electronic liquid crystals \cite{Sondhi-RMP,Fradkinbook,XGWenbook}.

Quantum critical points beyond the LGW paradigm have attracted increasing attentions. It is particularly intriguing to identify and understand quantum critical points which are forbidden according to the Landau criterion -- the so-called Landau-forbidden transitions. Remarkably, the theory of deconfined quantum critical points (DQCP) \cite{Senthil-04} provides an exotic scenario of realizing a continuous quantum phase transition between two symmetry-incompatible phases, which is putatively first-order according to the Landau symmetry criterion. Fractional excitations play an important role in such DQCP \cite{Senthil-04,Sandvik-2007, Melko-2008,Dunghai-2010,Shao-16}.

The Landau cubic criterion states that continuous phase transitions are also forbidden when cubic terms of order parameters are allowed by symmetry in the Landau-Ginzburg (LG) free energy. For instance, the quantum three-state Potts model in 2+1 or 3+1 dimensions has been convincingly shown to feature a first-order quantum phase transition \cite{Wu-82}, as cubic terms of the $Z_3$ order parameters are allowed and relevant in the low-energy LG free energy. One may naturally ask the following question: Is there any continuous transition that can violate this Landau criterion concerning cubic terms in LG free energy? 

\begin{figure}[t]				
\subfigure{\includegraphics[width=8.cm]{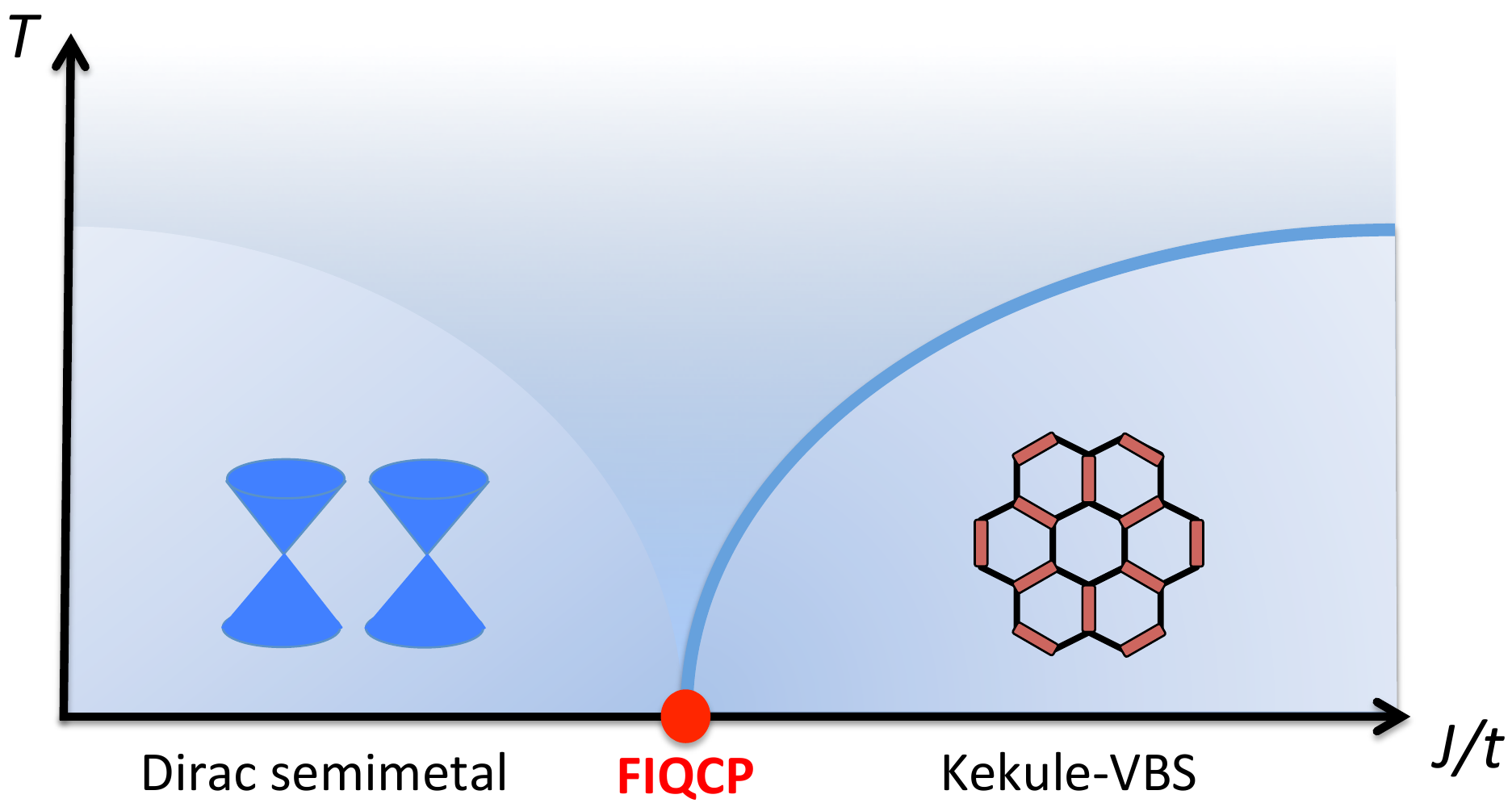}}
\caption{{$\mid$ \bf The fermion-induced quantum critical point (FIQCP).} According to the Landau cubic criterion, the transition would be putatively first-order because cubic terms of order-parameters are allowed by symmetry in the Landau-Ginzburg theory. However, it can be induced to be second-order by coupling to massless Dirac fermions. This FIQCP provides a new and generic scenario for transitions violating the Landau cubic criterion.}
\label{fig1}
\end{figure}

Here we discover an intriguing scenario violating the Landau cubic criterion; namely fermion-induced quantum critical points (FIQCP) are second-order quantum phase transitions induced by coupling gapless fermions to fluctuations of order parameters whose cubic terms appear in the Landau-Ginzburg theory. To be more explicit, we consider a quantum phase transition between Dirac semimetals in 2D \cite{Novoselov-05, Neto-09, Herbut-06,Assaad-13,Roy-13,Vafek-14} and Kekule valence bond solids (Kekule-VBS) \cite{Moessner-01,Chamon-2007,Ryu-2009,Pujari-13} with $Z_3$ symmetry-breaking where cubic terms are allowed in the LG free energy, as schematically shown in \Fig{fig1}. We perform RG analysis to show that this putative first-order phase transition can be driven to a continuous phase transition by fluctuations of gapless Dirac fermions. Our RG calculations are controlled by large-$N$ expansions where $N$ is the number of flavors of four-component Dirac fermions. Remarkably, the RG results identify a stable fixed point with vanishing cubic terms, which corresponds to a continuous phase transition between Dirac semimetals and Kekule-VBS, namely an FIQCP. To confirm the FIQCP obtained in the RG analysis, we consider microscopic models of $SU(N)$ fermions on the honeycomb lattice featuring the designed quantum phase transitions. No matter $N$ is even or odd, quantum Monte Carlo \cite{Scalpino-81,Fucito-81,Hirsch-81,Assaad-2005} simulations of these models can be made sign-problem-free by employing the Majorana method recently proposed by us in Ref. \cite{ZXLi-15a}. By large-scale sign-problem-free MQMC simulations, we show convincing evidences that the quantum phase transition between the Dirac semimetals and Kekule-VBS is continuous for $N=2, 3, 4, 5$, and $6$. The emergence of rotational symmetry at the transition reveals that this phase transition falls in chiral XY universality \cite{Rosenstein-93,Moshe-03}. We obtain various critical exponents at the FIQCP in MQMC simulations. Remarkably, the critical exponents derived from RG analysis reasonably agree with the ones obtained from our MQMC simulations, which strongly suggests that the FIQCPs
are robust. \\

\noindent {\bf Results}

\noindent {\bf Renormalization group analysis.} We begin by constructing the low-energy field theory describing the quantum phase transition. At low-energy and long-distance near the transition, the system can be described by Dirac fermions, fluctuating order parameters, and their couplings: $S=S_\psi+S_\phi+S_{\psi\phi}$. The action for Dirac fermions (on honeycomb lattice) is given by:
\bea
	S_\psi=\int d^3 x~ \psi^\dag [\partial_\tau -v(i\sigma^x \tau^z\partial_x+i\sigma^y\tau^0\partial_y)]\psi,
\eea
where $\tau^i$ ($\sigma^i$) Pauli matrices operate in valley (sublattice) space, $v$ denotes the Fermi velocity, and $\psi^\dag(x)=(\psi^\dag_{{\bf K}A}(x),\psi^\dag_{{\bf K}B}(x), \psi^\dag_{-{\bf K}A}(x),\psi^\dag_{-{\bf K}B}(x))$ is the four component fermion creation operator with $\pm {\bf K} = \pm(\frac{4\pi}{3},0)$ denoting valley momenta of Dirac points and $A,B$ labeling sublattices. Note that the spin index $\nu=1,\cdots,N$ is implicit in the action above; for spin-1/2 electrons in graphene $N=2$.

The Kekule-VBS order breaks lattice translational symmetry with wave vectors $\pm 2{\bf K}$ and also the $C_3$ rotational symmetry (see Supplementary Note 1 for details). The most general but symmetry constrained action describing the order-parameter fluctuations up to the fourth order is given by
\bea
S_\phi \!=\! \int\!\! d^3 x \Big[\!|\partial_\tau\phi|^2 \!+c^2 \! |\nabla \phi|^2 \!+\! r|\phi|^2\!+\! b(\phi^3+\phi^{\ast3})\!+\! u|\phi|^4\! \Big], ~~~\label{order_action}
\eea
where $\phi(x)\equiv \phi_{2\bf K}(x) $ is a complex order parameter, $c,r,b,u$ are real constants. According to the Landau criterion, the cubic terms above should render a first-order transition. Indeed, the action in Eq. \eqref{order_action} describes an effective field theory of quantum three-state Potts model which supports a weakly first-order quantum phase transition in 2+1 dimensions \cite{Wu-82,Pollmann-2015,Wu-2015}. However, as we shall show below, the coupling between the gapless Dirac fermions and order-parameter fluctuations will dramatically change this scenario by rendering the putative first-order transition into a continuous one. The Kekule-VBS ordering can gap out the Dirac fermions and the coupling between them reads
\bea
S_{\psi\phi} = g \int d^3 x(\phi \psi^\dag \sigma^x\tau^+ \psi + h.c.),
\eea
where $\tau^+=(\tau^x+i\tau^y)/2$ and $g$ labels the Yukawa-like coupling strength. Note that, although the fifth-order term $u_5(\phi^3+h.c.)|\phi|^2$ which is relevant at Gaussian fixed point in 2+1 dimensions and the sixth-order term $u_6|\phi|^6+u'_6(\phi^6+h.c.)$ which are marginal are allowed in the action by symmetry, these terms are irrelevant and can be safely omitted for $N>1/2$ at the FIQCP fixed point, as we shown explicitly in Supplementary Note 1.

\begin{figure}
\includegraphics[width=7.5cm]{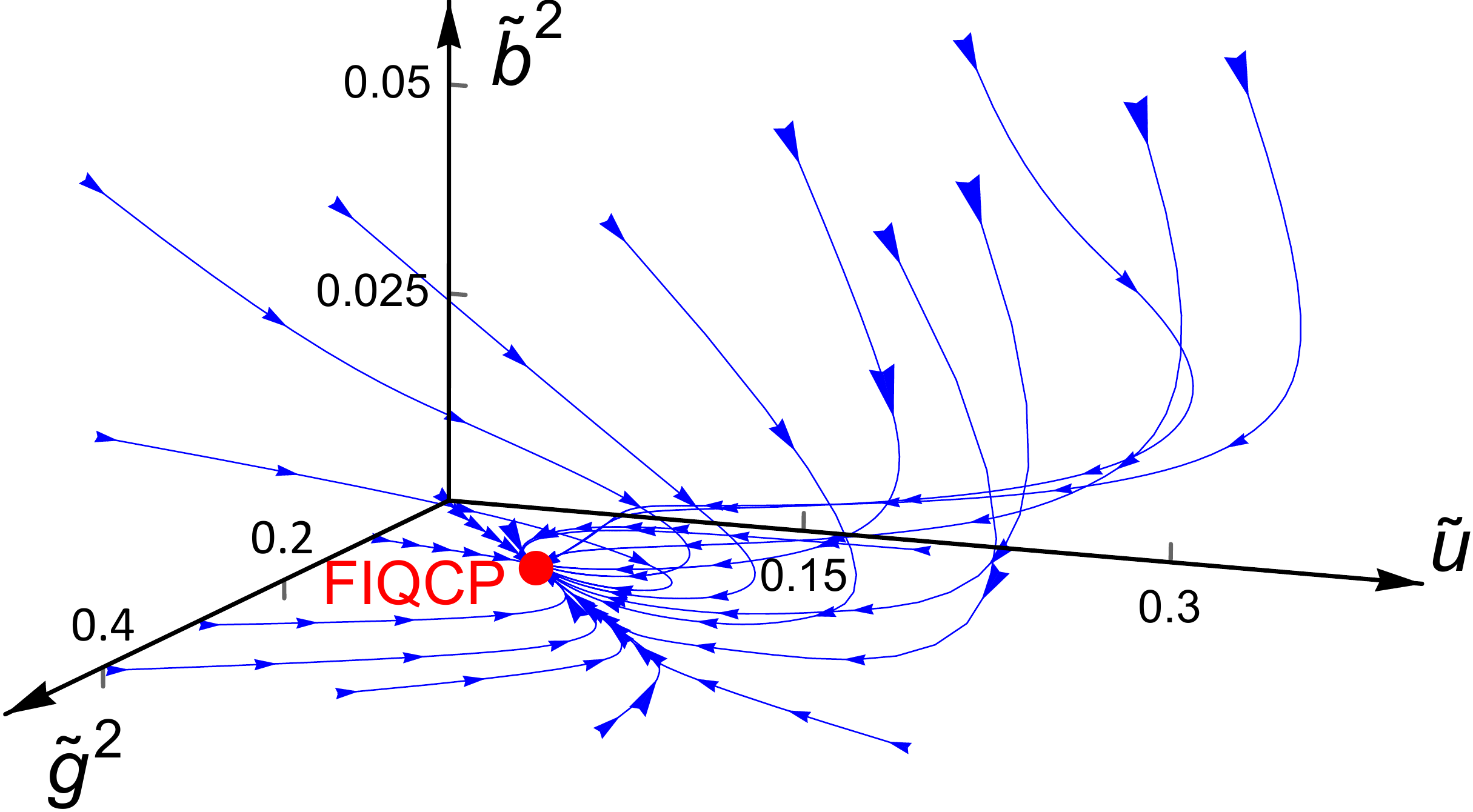}
\caption{{$\mid$ \bf The renormalization group flow of coupling constants.} From The only stable fixed point in the critical surface $r=r_c$ is denoted by the red point, which is the Gross-Neveu-Yukawa fixed point representing an FIQCP. Moreover, Lorentz symmetry emerges at the FIQCP as velocities of fermions and bosons flow to the same value. }\label{fig2}
\end{figure}

To answer whether the FIQCP occurs in the quantum phase transition, we perform RG analysis of the the effective field theory describing the phase transition. For simplicity, we employ dimensionless coupling constants $(\tilde{r},\tilde{g}^2, \tilde{b}^2,\tilde{u})$ (see Supplementary Note 1 for details). Integrating out the fast modes in the momentum-shell $\Lambda e^{-l} <p<\Lambda$ yields a set of RG equations, where $l>0$ parameterizes momentum-shell and $\Lambda$ is the ultraviolet cutoff. We implement a large-$N$ expansion in calculations where $N$ is number of fermion species ($N=2$ for spin-1/2 electrons in graphene). As long as $\tilde{g}^2$ stays non-zero at the infrared, the RG is controlled by a small parameter $1/N$. A second-order critical point should have only one relevant direction $\tilde{r}$, while those fixed points having more than one relevant directions are multi-critical points or indicate first-order transitions. In other words, a second-order critical point is stable under all perturbations in critical surface $ \tilde{r} \equiv \tilde{r}_c(\tilde{g}^2,\tilde{b}^2,\tilde{u})$, and, in particular, the cubic terms should be irrelevant.

By solving the RG equations, we find only one stable fixed point with $\tilde{g}^{\ast2} > 0$ on the critical surface, $(\tilde{g}^{\ast2}, \tilde{b}^{\ast2}, \tilde{u}^\ast)\approx (\frac{2}{\pi N},0,\frac{2}{\pi N})$, for $N>1/2$ (see Supplementary Note 1 for details). Moreover, the Fermi velocity and the boson velocity flow to the same value at the fixed point, indicating a Gross-Neveu-Yukawa (GNY) fixed point with emergent Lorentz symmetry. The flow of coupling constants near the stable GNY fixed point for $N=3$ is shown in \Fig{fig2}, where the red point denotes the GNY fixed point. In the vicinity of the GNY fixed point \cite{Herbut-06}, the linearized RG equations are given by (we set $v=c=1$ for simplicity)
\bea
	\frac{d \delta \tilde{g}^2}{dl} &=& - \delta \tilde{g}^2- \frac{9}{2 N} \delta \tilde{b}^2,  \\
	\frac{d \delta \tilde{b}^2}{dl} &=& - \frac{9}{N} \delta \tilde{b}^2, \\
	\frac{d \delta \tilde{u}}{dl} &=&  \frac{18}{N} \delta \tilde{g}^2+ \frac{99}{N} \delta \tilde{b}^2-\left(1+\frac{18}{N} \right) \delta \tilde{u},
\eea
from which it is obvious that the fixed point is a stable one as perturbations around the GNY point are irrelevant. The critical exponents at the GNY fixed point are given by $\eta=\frac{N}{N+1}, \nu^{-1} =2 -\frac{1+4N+ \sqrt{1+38 N+ N^2}}{5(1+N)}$. At the GNY fixed point, $\phi^5$ (also $\phi^6$) is irrelevant and can be safely neglected near the quantum phase transition for analyzing the FIQCP, as shown in the Supplementary Note 1.

The existence of the stable GNY fixed point in the large-$N$ RG analysis implies that the quantum phase transition is a continuous one with vanishing cubic terms and emergent rotational symmetry, namely an FIQCP for relatively large $N$. It is worth mentioning that, if $N=0$, the theory becomes a purely bosonic system which features a first-order phase transition as shown many years ago \cite{Wu-82}, consistent with the Landau cubic criterion. Moreover, for $N=1/2$, if the transition were be a continuous one, the critical point would feature an emergent spacetime supersymmetry (SUSY) \cite{Grover-14,SSLee-14,Jian-15}. Because of the emergent SUSY, the scaling dimension of the cubic term $\phi^3$ is known exactly to be 2, which is less than the spacetime dimension, implying that the cubic term is relevant at the SUSY fixed point and that it is in contradiction with the assumption that a continuous transition occurs for $N=1/2$. Consequently, the transition for $N=1/2$ must be first-order, which is an exact result! Our RG analysis predicts a critical $N_c$ ($N_c>1/2$) such that for $N>N_c$ the gapless fermions are able to drive such a putative first-order transition into a continuous one, and that for $N<N_c$ the putative first-order transition survives. As the RG analysis is controlled by the $1/N$ expansion, determining the exact value of such critical $N_c$ is beyond the RG scheme here. Sign-problem-free MQMC simulations below give an upper bound of $N_c$, namely $N_c<2$, as the simulations convincingly show that the FIQCP occurs for $N=2,3,4,5,6$.\\

\begin{figure}[t]
\includegraphics[width=6.0cm]{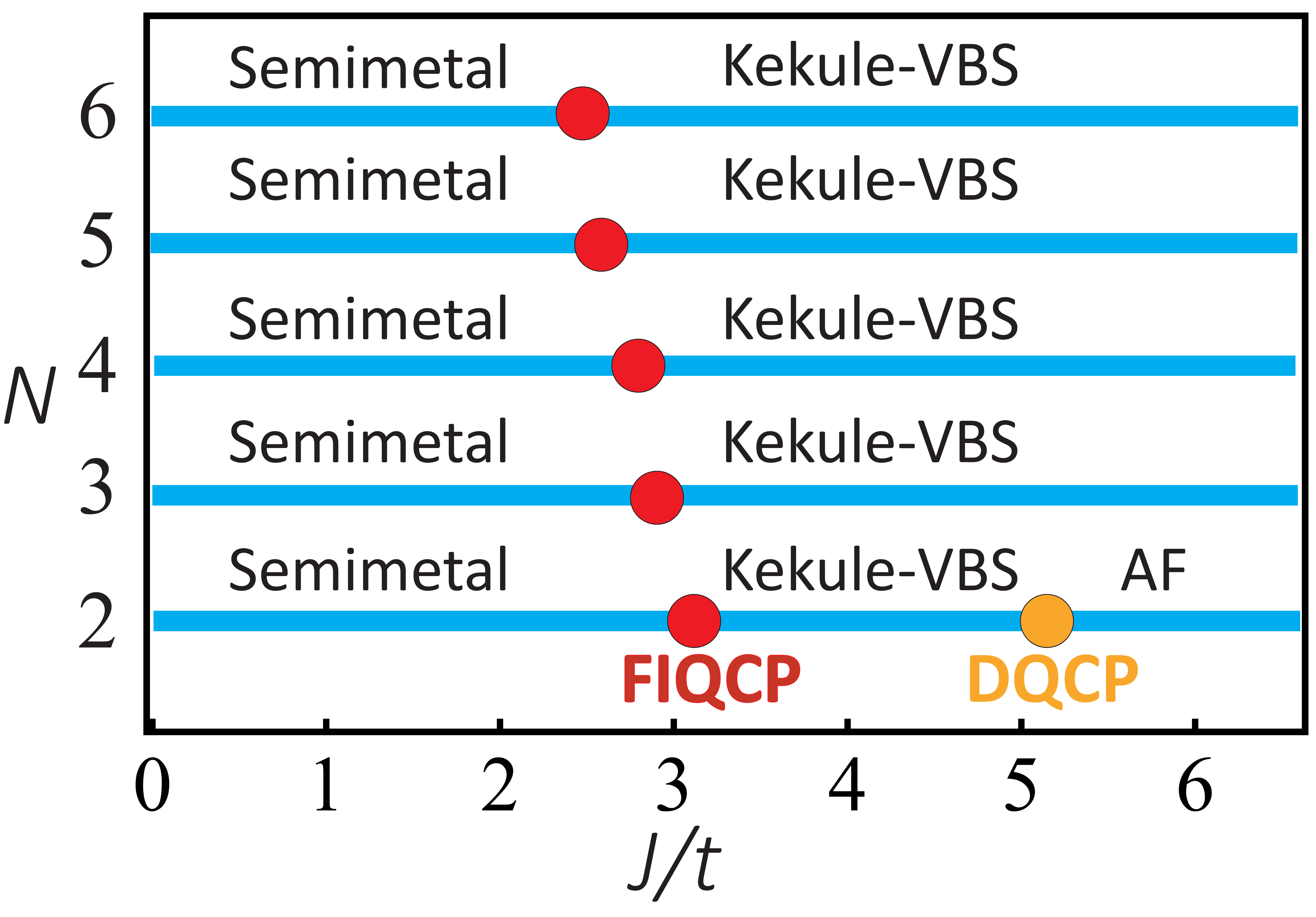}
\caption{{$\mid$ \bf The quantum phase diagram obtained from QMC simulations.} The quantum phase diagram of the models of SU($N$) fermions on the honeycomb lattice, for $N=2,3,4,5,6$, is obtained by sign-problem-free Majorana quantum Monte Carlo (QMC) simulations. For $N =2,3,4,5,6$, the system encounters a continuous transition violating the Landau cubic criterion, dubbed as fermion-induced quantum critical point (FIQCP) here, between the Dirac semimetals to the Kekule valance bond solid (Kekule-VBS) phase. For $N=2$, the quantum phase transition between the Kekule-VBS and the antiferromagnetic (AF) phases is a deconfined quantum critical point (DQCP).}
\label{fig3}
\end{figure}

\noindent{\bf Majorana quantum Monte Carlo simulations.} In order to confirm the scenario of FIQCP obtained in the RG analysis above, we introduce a sign-problem-free model of SU($N$) fermions \cite{Affleck-1988,Sachdev-1989,Wu-2003,Honerkamp-2004,Troyer-2011,Congjun-2013,Assaad-2013,Block-13} on the honeycomb lattice, which features a quantum phase transition between Dirac semimetals and the Kekule-VBS phase, as follows:
\bea
H= -t\sum_{\avg{ij}}\big[c_{i\alpha}^{\dagger}c_{j\alpha} + h.c.\big]\!-\! \frac{J}{2N}\sum_{\avg{ij}}\big[ c^{\dagger}_{i\alpha} c_{j\alpha}+ h.c.\big]^2,  ~~
\label{model}
\eea
where the summation over spin species $\alpha=1,\cdots,N$ is implicitly assumed,  $c^\dagger_{i\alpha}$ is the creation operator of fermions with spin index $\alpha$ on site $i$,  $t$ is the hopping amplitude, and $J$ is the strength of interactions. Hereafter we set $t=1$ as the energy unit. The low-energy physics of noninteracting $SU(N)$ fermions at half-filling in Eq. \eqref{model} can be described by $N$ massless four-component Dirac fermions. The system can undergo a quantum phase transition from Dirac semimetals to the Kekule-VBS phase as the interaction $J$ is increased. Most remarkably, no matter $N$ is odd or even, this model is free from infamous fermion-sign-problem \cite{Loh-1990,Wu-2005,Troyer-2005,Berg-2012,Lei-2015,ZXLi-2016sign,TXiang-2016sign} when the Majorana representation \cite{ZXLi-15a} is used, which allows us to do unbiased simulations to investigate the nature of this quantum phase transition in systems with large lattice sizes.

As we are interested in quantum phase transitions, we use projector QMC \cite{Sorella-1989,White-1989} to explore ground state properties of the model in \Eq{model}.  To study the transition into the Kekule-VBS phase, we calculate the structure factor of VBS order parameters by MQMC: $S_{\text{VBS}}({\bf k}, L) = \frac{1}{L^4}\sum_{i,j} e^{i{\bf k}\cdot(\bf{r_i}-\bf{r_j})} \langle (c^{\dagger}_i c_{i+\delta} + h.c.)(c^{\dagger}_j c_{j+ \delta} + h.c.)\rangle$, where the system has $2\times L\times L$ sites with periodic boundary condition and $\delta$ labels the direction of a nearest-neighbor bond; the Kekule-VBS order parameter $\Delta_{\text{VBS}}$ can be obtained through $\Delta_{\text{VBS}}^2 = \lim_{L\rightarrow \infty } S_{\text{VBS}}({\bf K},L)$, where ${\bf K}$ is the VBS ordering vector. It is a finite value when the system lies in the Kekule-VBS phase.

\begin{figure}[t]
\includegraphics[width=8.4cm]{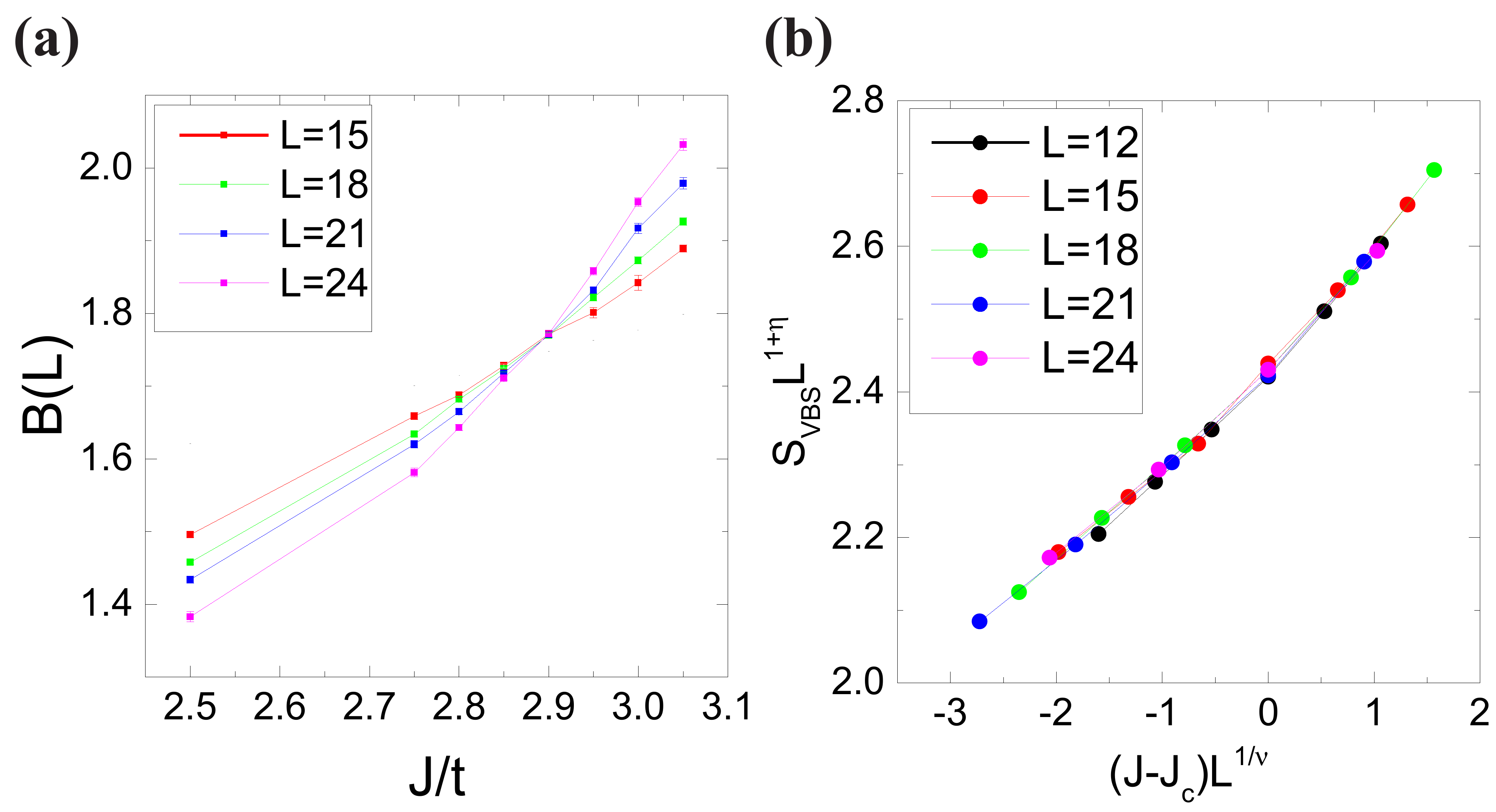}
\caption{ {$\mid$ \bf The Binder ratio and data collapse analysis.} ({\bf a}) The Binder ratio with different $J/t$ and different $L$ up to $L=24$ for $N=3$ is obtained from sign-problem-free Majorana QMC; the Kekule-VBS phase transition occurs at $J_c/t \approx 2.9$.  ({\bf b}) Data collapsing according to the scaling relation is used to determine $\nu$. From the fitting, $\nu = 1.07$ for $N=3$. }
\label{fig4}
\end{figure}

We perform large-scale MQMC simulations for $N = 2,3,4,5,6$ with $L=12,15,18,21,24$. For simplicity, we shall show $N=3$ results here and the details about other $N$ can be seen in Supplementary Note 2. As shown in \Fig{fig4}(a), the critical value $J_c$ can be obtained through the Binder ratio \cite{Assaad-2005} of Kekule-VBS order: $B(L)\equiv \frac{S_{\text{VBS}}({\bf K},L)}{S_{\text{VBS}}({\bf K}+ \delta {\bf k},L)}$, where $|\delta {\bf k}|=\frac{2\pi}{L}$. At the putative critical point, the Binder ratios of different $L$ should cross at the same point for sufficiently large $L$. From MQMC simulations, we calculated the Binder ratio around the Kekule-VBS phase transitions and obtained $J_c$ for $N=2,3,4,5$ and 6, as shown in \Fig{fig3}. The case of $N=3$ is shown in \Fig{fig4}(a) where $J_c \approx 2.9t$, while the results for other $N$ are shown in Supplementary Figure 1.

To better answer the question whether the Kekule-VBS transition is first-order or continuous, we further investigate the critical behaviour around the transition. Two independent critical exponents, $\eta$ and $\nu$, can be obtained by MQMC simulations and other critical exponents such as $\beta$ may be obtained from $\eta$ and $\nu$ through hype-scaling relations \cite{Sorella2016}. The critical exponents $\eta$ and $\nu$ satisfy the following scaling relation: $S_{\text{VBS}}({\bf K},L) = L^{-z-\eta} {\cal F}( L^{1/\nu}( J - J_c ))$ for $J$ close to $J_c$ and relatively large $L$; we assume the dynamical exponent $z=1$ mainly because of massless Dirac fermions. First, we obtain $\eta$ by plotting $S_{\text{VBS}}({\bf K},L)$ at $J=J_c$ versus $L$ in a log-log way and then fitting it to a linear function with slope $-(1+\eta)$, as shown in Supplementary Figure 2. Second, there exists an appropriate value of $\nu$ such that different points ($S_{\text{VBS}}({\bf K},L) L^{1+\eta}, L^{1/\nu}(J-J_c)$) of different $J$ around $J_c$ and different $L$ should collapse on a single unknown curve ${\cal F}$, as shown in \Fig{fig4}(b). For $N=3$, such finite-size scaling analysis gives rise to $\eta \approx 0.78$ and $\nu \approx 1.07$ as shown in \Fig{fig4}. The finite-size scaling analysis for other $N$ is shown in Supplementary Figure 3.

\begin{table}[t]
\centering
\caption{{ \bf The critical exponents}. $\eta$ and $\nu$ at the FIQCPs obtained in the present work by QMC and RG (large-$N$), respectively, for $N$ = 2, 3, 4, 5, and 6. The comparisons with Ref. \cite{Rosenstein-93} are also shown.} \label{comparison}
\begin{tabular}{|c|c|c|c|}
\hline
 ~~~~~$N$~~~~~& ~~~~~~~~~~~~Method~~~~~~~~~~~~ & ~~~~~~$\eta$~~~~~~& ~~~~~~$\nu$~~~~~~   \\
\hline
 & large-$N$ (present) & $0.67$ & $1.25$    \\
\cline{2-4}
$2$ & QMC (present) & $0.71(3)$ & $1.06(5)$    \\
\cline{2-4}
 & $4-\epsilon$, two-loop \cite{Rosenstein-93} & $0.67$ & $0.94$    \\
\hline
 & large-$N$ (present) & $0.75$ & $1.26$    \\
\cline{2-4}
$3$ & QMC (present) & $0.78(2)$ & $1.07(4)$    \\
\cline{2-4}
 & $4-\epsilon$, two-loop \cite{Rosenstein-93} & $0.77$ & $0.96$    \\
\hline
 & large-$N$ (present) & $0.80$ & $1.25$    \\
\cline{2-4}
$4$ & QMC (present) & $0.80(4)$ & $1.11(3)$    \\
\cline{2-4}
 & $4-\epsilon$, two-loop \cite{Rosenstein-93} & $0.82$ & $0.97$    \\
\hline
 & large-$N$ (present) & $0.83$ & $1.23$    \\
\cline{2-4}
$5$ & QMC (present) & $0.85(4)$ & $1.07(2)$    \\
\cline{2-4}
 & $4-\epsilon$, two-loop \cite{Rosenstein-93} & $0.86$ & $0.97$    \\
\hline
 & large-$N$ (present) & $0.86$ & $1.22$    \\
\cline{2-4}
$6$ & QMC (present) & $0.87(4)$ & $1.06(3)$    \\
\cline{2-4}
 & $4-\epsilon$, two-loop \cite{Rosenstein-93} & $0.88$ & $0.98$    \\
\hline
\end{tabular}
\end{table}

We summarize the results of $\eta$ and $\nu$ for $N = 2,3,4,5,6$ obtained from QMC simulations and RG analysis, respectively, in Table \ref{comparison}.
One can see that the values of $\eta$ obtained by QMC and RG are in very good agreement with each other. The agreement in $\nu$ is not as good as $\eta$, but should become better for larger $N$ as the RG analysis is performed in large-$N$ expansion.  The reasonable agreement between RG and QMC results for $N=2,3,4,5,6$ convincingly suggests that the quantum phase transition between the Dirac semimetals and the Kekule-VBS phases for $N\ge 2$ is an FIQCP. The irrelevance of the cubic terms of the VBS order parameter at the FIQCP is further evidenced by the emergence of order-parameter U(1) symmetry at the quantum phase transition, as shown in the Supplementary Figure 4. \\

\noindent{\bf Discussion}

\noindent One possible material candidate to realize FIQCP is graphene. Because of its spin-1/2 degree of freedom, graphene hosts $N=2$ Dirac fermions. It would be intriguing to observe FIQCPs in graphene-like systems. Indeed, it was reported that Kekule-VBS ordering has been experimentally observed under certain conditions in graphene \cite{columbia}. According to our RG analysis and MQMC simulations, we predict that the quantum phase transitions into Kekule-VBS in the $N=2$ Dirac systems (i.e. graphene-like materials) are FIQCPs. Other materials which might feature an FIQCP include 2H-TaSe$_2$, for which the $3\times 3$ charge-density wave ordering seems to be a continuous transition \cite{Moncton-1977} even though according to Landau it would be first-order.

We have shown that a putative first-order phase transition can be driven to a continuous one by coupling to $N$ massless Dirac fermions in 2+1 dimensions, from both sign-problem-free Majorana QMC simulations with $N\ge 2$ and the large-$N$ RG analysis. To the best of our knowledge, it is for the first time that an FIQCP in 2+1 dimensions is convincingly established as a quantum phase transition violating the Landau cubic criterion. Our proposal of FIQCP in the present paper was further confirmed by subsequent works which find interesting examples of FIQCP in other types of theories \cite{Herbut2016,Jian2016}. We would also like to mention that our simulations show evidences of a deconfined quantum critical point (DQCP) between the Kekule-VBS and antiferromagnetic (AF) phases occurring in the $N$=$2$ model in \Eq{model}, as shown in \Fig{fig3}. We believe that our study has provided new insights towards a unified understanding of quantum critical points and shall pave a new avenue to understand exotic quantum phase transitions beyond the conventional LGW paradigm \cite{Cenke-2012}. \\

\noindent{\bf Methods} \\
{\footnotesize \noindent{\bf Renormalization group analysis.} The renormalization group equations are obtained by one-loop large-$N$ calculations in 2+1 dimensions.  \\
{\bf Quantum Monte Carlo.} The projector QMC calculation is carried out to explore the ground-state properties of model \Eq{model}. In projector QMC, the ground state is obtained through projection: $\ket{\psi_0}= \lim_{\Theta \rightarrow \infty } e^{-\Theta H} \ket{ \psi_{T}}$,  where $ \ket{\psi_{T}}$ is a trial wave function and is chosen to be the ground state of the non-interacting part of in \Eq{model}. Because of the absence of sign-problem in Majorana representation, we can perform large-scale QMC simulations with large system sizes and sufficiently large $\Theta$. In our simulations, we use $\Theta =40$ and also have checked that results stay nearly the same for larger $\Theta$. We set discrete imaginary-time step $\Delta \tau = 0.04$ and verified that the results do not change within our statistics errors if we use smaller $\Delta \tau$. Simulations are performed in honeycomb lattice of linear sizes $L= 12,15,18,21,24$ with $N = 2 L^2$ sites. Binder ratio and data collapse techniques are implemented to identify the accurate QCP and extract critical exponents.\\
{\bf Data availability.} The data that support the findings of this study are available from the corresponding author (H.Y.) upon request.} \\

\noindent{\bf Acknowledgement}\\
{\footnotesize \noindent We sincerely thank Cenke Xu and Shou-Cheng Zhang for helpful discussions. This work is supported in part by the National Natural Science Foundation of China under Grant No. 11474175 (Z.-X.L., S.-K.J., Y.-F.J., H.Y.), by the Ministry of Science and Technology of China under Grant No. 2016YFA0301001 (H.Y.), and by the National Thousand-Young Talents Program (H.Y.). \\
\noindent $^\ast$Correspondence and requests for materials should be addressed to H.Y. (yaohong@tsinghua.edu.cn). Z.-X.L., Y.-F.J., and S.-K.J. contributed equally to this work. 
}

\begin{widetext}
\section{Supplementary Information}
\renewcommand{\theequation}{S\arabic{equation}}
\setcounter{equation}{0}
\renewcommand{\thefigure}{S\arabic{figure}}
\setcounter{figure}{0}

{\bf Supplementary Note 1: Renormalization group analysis} \\

 As mentioned in the main text, at the quantum phase transition point, the low-energy and long-distance physics can be described by the massless Dirac fermions, the fluctuating Kekule valance bond solid (Kekule-VBS) order parameters, and the couplings between fermions and order parameter fields: $S=S_\psi+S_\phi+S_{\psi\phi}$. The action for massless Dirac fermions on the honeycomb lattice is given by:
\bea
	S_\psi=\int \frac{d^d k}{(2\pi)^d} \psi^\dag [-iw+v(k_x\sigma^x \tau^z+k_y \sigma^y\tau^0)] \psi,
\eea
where $\tau^i$ $(\sigma^i)$ are Pauli matrices with valley (sublattice) indices, and $\psi^\dag=(\psi^\dag_{{\bf K}A},\psi^\dag_{{\bf K}B}, \psi^\dag_{-{\bf K}A},\psi^\dag_{-{\bf K}B})$ is fermion creation operators around the Dirac points $\pm {\bf K}=\pm (\frac{4\pi}{3},0)$.

The Kekule-VBS order parameter possesses $2{\bf K}$ momentum and serves a two-dimensional irreducible representation of the $C_3$ group, which is generated by translation operators $T_1,T_2 \equiv T_1^2$, where $T_{1}$ denotes translating by lattice vector ${\bf e_1}=(1,0)$ and $T_2$ denoting ${\bf e_2}=(\frac{1}{2},\frac{\sqrt{3}}{2})$. Static Kekule-VBS ordering can generate a mass for Dirac fermions. Thus, the Kekule-VBS order parameters is given by $\phi_{2{\bf K}} = \langle \psi_{\bf K}^\dag \sigma^x \psi_{-\bf K} \rangle$.  The transformation laws for the Kekule-VBS order parameters read:
\bea
	T_1(\phi_{2{\bf K}})&=& e^{i \frac{4\pi}{3}}\phi_{2{\bf K}}, \\
    T_2(\phi_{2{\bf K}}) &=& e^{i \frac{2\pi}{3}}\phi_{2{\bf K}},\\
    P_x(\phi_{2{\bf K}})&=& \phi^\ast_{2{\bf K}},\\
    P_y(\phi_{2{\bf K}})&=&\phi_{2{\bf K}},
\eea
where $P_i$ is the reflection operator sending $i$ to $-i$. The most general action for the Kekule-VBS order parameter allowed by the symmetries up to fourth-order is given by:
\bea
	S_\phi=\int d^d x \Big[ |\partial_\tau\phi|^2+c^2|\nabla\phi|^2+ r|\phi|^2+ b(\phi^3+\phi^{\ast3})+ u|\phi|^4 \Big],
\eea
where $\phi\equiv \phi_{2{\bf K}}$ for simplicity. The cubic terms in the action are allowed by $C_3$ symmetry and $b$ is a real constant as required by the reflection symmetries.
Dictated by symmetry, the most relevant fermion-boson coupling reads:
\bea
	S_{\psi\phi} &=&g \int d^d x (\phi_{2\bf{K}} \psi^\dag_{\bf{K}} \sigma^x \psi_{-\bf{K}} + h.c.), \\
&=& g \int d^d x (\phi \psi^\dag \sigma^x \tau^+ \psi + \phi^\ast \psi^\dag \sigma^x \tau^- \psi),
\eea
where $\tau^\pm \equiv \frac{1}{2}( \tau^x \pm \tau^y)$ and we have used the gauge freedom to fix $g$ to be real. For convenience, we define gamma matrix: $\gamma^0=\sigma^z,\gamma^1=-\sigma^y\tau^z,\gamma^2=\sigma^x,\gamma^3=\sigma^y\tau^x,\gamma^5=\sigma^y\tau^y$, which have the properties $\{ \gamma^\mu,\gamma^\nu \}=2\delta^{\mu\nu}$, and  $\gamma^\pm=\frac{1}{2}(\gamma^3 \pm i\gamma^5)$. In the following, we use Greek letter to denote $0,1,2$ and English letter to denote $\pm$. In this convention, the action is given by $S=S_\psi+S_{\phi,0}+S_\textrm{int}$ with
\bea
	S_\psi &=&  \int \frac{d^d k}{(2\pi)^d} \bar{\psi} (-i\gamma^0 k_0-i v \gamma^i k_i) \psi, \\
	S_{\phi,0} &=&  \int \frac{d^d k}{(2\pi)^d} \phi^\ast(k_0^2+c^2 k_i^2+r) \phi, \\
	S_\textrm{int} &=&  \int d^d x \left[  b(\phi^3+\phi^{\ast3}) + u|\phi|^4+ ig( \phi \bar{\psi} \gamma^+ \psi +\phi^\ast \bar{\psi} \gamma^- \psi )\right].
\eea	
where $\bar{\psi}= \psi^\dag \gamma^0$ and $k_\mu=(w,k_x,k_y,k_z)$ and $k^2=k_\mu k_\nu \delta^{\mu\nu}$. The Feynman propagators are given by
\bea
	S(k) &=& \frac{i (\gamma^0 k_0+ v \gamma^i k_i)}{k_0^2+ v^2 k_i^2},\\
 D(k) &=& \frac{1}{k_0^2+c^2 k_i^2+r }.
\eea
In the following calculation, the gamma matrices have the properties: $\{\gamma^i,\gamma^j\}=2g^{ij},\{\gamma^i,\gamma^\mu\}=0,\{ \gamma^\mu,\gamma^\nu \}=2\delta^{\mu\nu}$, where $g^{ij}=(1-\delta^{ij})/2$. Moreover, the trace properties are listed here for reference:
\bea
\text{Tr}[\gamma^i\gamma^\mu\gamma^j\gamma^\nu] &=& -4N g^{ij} \delta^{\mu\nu}, \\
\text{Tr}[\gamma^\mu \gamma^\nu\gamma^\rho \gamma^\sigma] &=& 4N[\delta^{\mu\nu}\delta^{\rho\sigma}-\delta^{\mu\rho}\delta^{\nu\sigma}+\delta^{\mu\sigma}\delta^{\nu\rho}], \\
\text{Tr}[\gamma^i\gamma^\mu\gamma^j\gamma^\nu\gamma^k\gamma^\rho] &=& 0,\\
\text{Tr}[\gamma^i\gamma^\mu\gamma^j\gamma^\nu\gamma^k\gamma^\rho\gamma^l\gamma^\sigma] &=& 4N[\delta^{\mu\nu}\delta^{\rho\sigma}-\delta^{\mu\rho}\delta^{\nu\sigma}+\delta^{\mu\sigma}\delta^{\nu\rho}][g^{ij}g^{kl}-g^{ik}g^{jl}+g^{il}g^{jk}],
\eea
where we have promoted the spin $SU(2)$ to $SU(N)$. We implement a large-$N$ expansion in calculation. We will see that as long as $g^2$ stays non-zero at the infrared, the renormalization group (RG) procedure is controlled by $1/N$. This means the critical $\tilde{r}=\tilde{r}_c$ is of order $1/N$; consequently setting $\tilde{r}_c=0$ in calculations will not affect the result in the lowest order. We use a momentum-shell renormalization scheme. Namely, the fast modes in the spherical momentum-shell $\Lambda e^{-l} < p <\Lambda$ are integrated out, giving renormalization to the slow modes with $p<\Lambda e^{-l}$, where $l>0$ is the flow parameter. To accommodate this renormalization effects, various coupling constants begin to run when energy scale changing. The relevant renormalization comes from $S_{\text{eff}}= S_<- \frac{1}{2} \langle S^2_c \rangle_>+ \frac{1}{6} \langle S^3_c \rangle_>-  \frac{1}{24} \langle S^4_c \rangle_>$ to one-loop level, where $>,<$ denote the fast and slow modes, respectively, and $\langle \cdots \rangle_>$ means taking expectation value in fast mode configurations. Boson self-energy $\Pi(p)$ and fermion self-energy $\Sigma(p)$ are
\bea
\Pi(p) &=& \int \frac{d^dk}{(2\pi)^d} \text{Tr}[(ig\gamma^+)S(p+k)(ig\gamma^-)S(k)]-18b^2 \int \frac{d^dk}{(2\pi)^d} D(p+k)D(k), \nn \\
 &=& g^2 \frac{N \pi }{2v^3} K_d \Lambda^{d-4}l  (p_0^2+ v^2 p_i^2)+ b^2 \frac{9\pi}{4c^5} K_d\Lambda^{d-6}l (p_0^2-\frac{1}{3}c^2 p_i^2 ) ,
\eea
and
\bea
\Sigma(p) &=& -2 \int \frac{d^dk}{(2\pi)^d} (ig\gamma^-)S(p+k)(ig\gamma^+)D(k), \\
&=&g^2 \frac{\pi}{c(c+v)^2} K_d\Lambda^{d-4}l [\gamma^0(-ip_0)+\frac{2c+v}{3v}\gamma^i (-ivp_i)],
\eea
where $K_d=\frac{A_{d-1}}{(2\pi)^d}$, and $A_{d}$ is the surface area of unit $d$-sphere. Renormalization contribution to three-boson-vertex is
\bea
\Gamma_{\phi^3} =
\Gamma_{\phi^{\ast3}} &=& -6b u \int \frac{d^d k}{(2\pi)^d} D^2(k)-\frac{1}{3}\int \frac{d^d k}{(2\pi)^d} \text{Tr}[ S(k)(ig\gamma^-)S(k)(ig\gamma^-)S(k)(ig\gamma^-)],\\
&=&-b u  \frac{3\pi}{c^3} K_d \Lambda^{d-4}l,
\eea
and to four-boson-vertex is
\bea
\Gamma_{|\phi|^4} &=& -10u^2\int \frac{d^d k}{(2\pi)^d} D^2(k)+144 u b^2 \int \frac{d^d k}{(2\pi)^d} D^3(k)-324b^4\int \frac{d^d k}{(2\pi)^d} D^4(k) \nn\\
&& +\int \frac{d^d k}{(2\pi)^d} \text{Tr}[S(k)(ig\gamma^+)S(k)(ig\gamma^-)S(k)(ig\gamma^+)S(k)(ig\gamma^-)] , \nn\\
&=&- u^2 \frac{5\pi}{c^3} K_d \Lambda^{d-4} l +  u b^2\frac{54\pi}{c^5} K_d \Lambda^{d-6} l + g^4 \frac{N \pi}{2v^3} K_d\Lambda^{d-4} l   -b^4\frac{405 \pi}{4 c^7} K_d \Lambda^{d-8} l .
\eea
The contribution from integrating out the fast modes to fermion-boson-vertex vanishes because of properties of $\gamma^\pm$ structure. After rescaling $p \rightarrow e^{-l} p$, we bring the renormalization effects into the flow of coupling constants. This results in the following RG equations:
\bea
&&	\frac{dc}{dl} = - \frac{N\pi(c^2-v^2)}{4 c^3 v^3}\tilde{g}^2 - \frac{3\pi}{2 c^6} \tilde{b}^2, \\
&& \frac{dv}{dl} = - \frac{2 \pi(v-c)}{3cv(c+v)^2} \tilde{g}^2, \\
&& \frac{d \tilde{g}^2}{dl}=(4-d) \tilde{g}^2- \frac{9\pi}{4c^5} \tilde{b}^2 \tilde{g}^2- \left(\frac{N\pi}{2v^3}+ \frac{2\pi}{c(c+v)^2} \right) \tilde{g}^4, \\
&& \frac{d \tilde{b}^2}{dl}= (6-d)\tilde{b}^2 -\frac{3N\pi}{2v^3} \tilde{g}^2 \tilde{b}^2-\frac{6\pi}{c^3} \tilde{b}^2 \tilde{u}- \frac{27\pi}{4c^5} \tilde{b}^4 ,\\
&& \frac{d \tilde{u}}{dl}=(4-d)\tilde{u}-\frac{N\pi}{v^3} \tilde{g}^2 \tilde{u} +\frac{N \pi}{2v^3} \tilde{g}^4- \frac{5\pi}{c^3} \tilde{u}^2+\frac{99\pi}{2 c^5}  \tilde{u}\tilde{b}^2   -\frac{405\pi}{4c^7}\tilde{b}^4 ,
\eea
where we define dimensionless coupling constants: $\tilde{g}^2= K_d \Lambda^{d-4} g^2,\tilde{b}^2= K_d \Lambda^{d-6} b^2, \tilde{u} = K_d \Lambda^{d-4} u$.
Solving these coupled RG equations in (2+1)-dimensions ($d=3$), we find four fixed points totally. On one hand, two of them locates in the ($\tilde b^2$,$\tilde u$) plane with $\tilde{g}^2=0$ and are the familiar Gaussian ($\tilde b^{\ast 2}=0$ and $\tilde u^\ast =0$) and Wilson-Fisher fixed points ($\tilde b^{\ast 2}=0$ and $\tilde u^\ast=1/(5\pi)$), respectively. Both fixed points are unstable. Since the fermions decouple with order parameters at these fixed point, they are not controlled by $1/N$ and are not relevant in this work.

On the other hand, there are two fixed points with non-zero fermion-boson coupling. But, only one of them is a physically meaningful fixed point with $u^\ast>0$ while the other one with $u^\ast<0$ is not physically meaningful as the free energy is not bounded from below. At the physical fixed point, the fermion and the boson velocities flow to the same value, i.e., $c^\ast=v^\ast$. Thus, we set $c^\ast=v^\ast=1$ below for simplicity. The coupling constants at this fixed point are given by $(\tilde{g}^{\ast2}, \tilde{b}^{\ast2}, \tilde{u}^\ast) =(\frac{2}{\pi(N+1)}, 0, R)$, where $R=\frac{1-N+\sqrt{N^2+38N+1}}{10\pi(N+1)}$. Remarkably, the rotational and Lorentz symmetries emerge at low-energy and long-distance as $c^\ast=v^\ast$ and $\tilde{b}^{\ast2}=0$, which is a Gross-Neveu-Yukawa (GNY) fixed point corresponding to chiral XY universality. This GNY fixed point is stable for relatively large $N_c$ as indicated by the flow diagram as well as the linearized RG equations given in main text and controls the behaviors at the FIQCP. The scaling fields near this fixed point read $(-1, -\frac{\sqrt{N^2+38N+1}}{N+1}, \frac{3(N+4- \sqrt{N^2+38N+1})}{5(N+1)} )$, from which one can find the critical number reads $N_c=\frac{1}{2}$. For $N>N_c$, the GNY fixed point is stable due to the negativity of all scaling fields.

\begin{figure*}
\includegraphics[width=17.4cm]{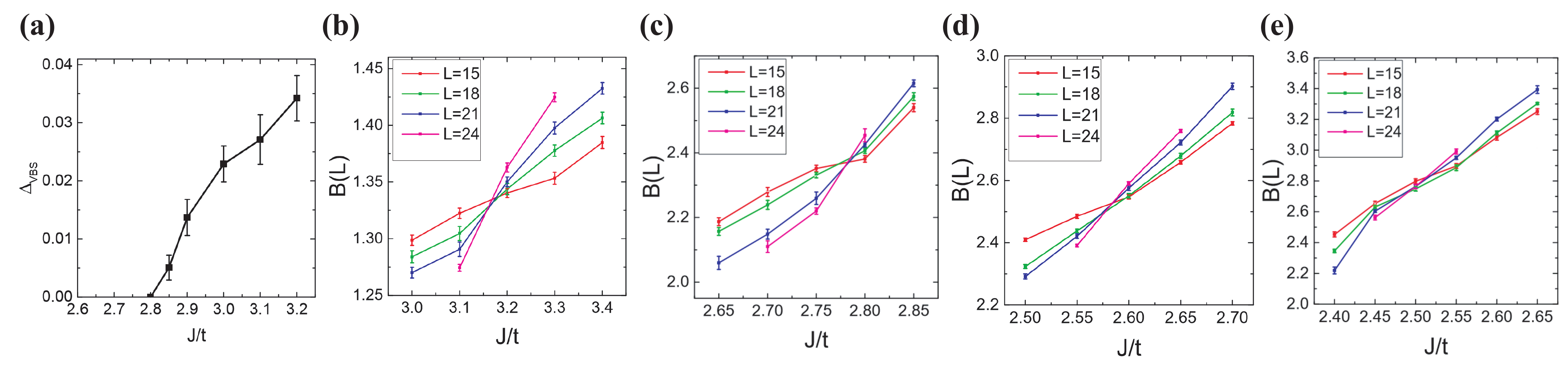}
\caption{ {$\mid$ \bf The Kekule-VBS order parameter and Binder ratio.} ({\bf a}) The Kekule valance bond solid (Kekule-VBS) order parameter versus $J/t$ for $N=4$. The Kekele-VBS order parameter is extracted by fitting VBS structure factor using second-polynomials in $1/L$. The standard method of least square is employed to fit the second-polynomials function of $1/L$. The error bar denotes the standard error of the intercept of second-polynomials function. ({\bf b}-{\bf e}) The Binder ratio results for different $N$: ({\bf b}) $N=2$, ({\bf c}) $N$=4, ({\bf d}) $N$=5, ({\bf e}) $N=6$; The phase transition point of $J/t$ is the crossed point of Binder ratio curve with different $L$. \\}
\label{figs1}
\end{figure*}

We believe that the second-order phase transition found in Majorana quantum Monte Carlo (MQMC) simulation is controlled by the GNY fixed point. At this fixed point fermion and boson fields will get non-trivial anomalous dimensions:
\bea
	\eta_\phi &=& \frac{N\pi}{4} \tilde{g}^{\ast2}+ \frac{9\pi}{8}\tilde{b}^{\ast2}=\frac{N}{2N+2},\\
\eta_\psi &=&\frac{\pi}{8} \tilde{g}^{\ast2} = \frac{1}{4(N+1)}.
\eea
The critical exponent $\eta$ is directly related to $\eta \equiv 2\eta_\phi=\frac{N}{N+1}$. To determine other critical exponents, we calculate the $\phi^2(x)$ vertex:
\bea
	\Gamma_{|\phi|^2}&=&-18b^2 \int \frac{d^dk}{(2\pi)^d} \frac{1}{(k^2+r)^2}+ 4u  \int \frac{d^dk}{(2\pi)^d} \frac{1}{k^2+r},\\
&=&\left(- \frac{9\pi\tilde{b}^2}{(1+\tilde{r})^{3/2}} + \frac{4\pi\tilde{u}}{\sqrt{1+\tilde{r}}} \right) \Lambda^2 l ,
\eea
where $\tilde{r} = \Lambda^{-2} r$. From this contribution, we have $\eta_r=-\eta+ \frac{27\pi}{2} \tilde{b}^{\ast2}-2\pi\tilde{u}^\ast=-\frac{1+4N+ \sqrt{1+38 N+ N^2}}{5(1+N)}$. Thus, the critical exponent at the GNY fixed point is given by $\nu^{-1}=2+ \eta_r $ which gives rise to the formula in the main text.

We consider the insertion of the term $u_5 (\phi^3+h.c.)|\phi|^2$ into the fixed-point action and ask whether it would affect the GNY fixed point or not. The RG equation of $u_5$ is given by
\bea
\frac{du_5}{dl}=\Big[\frac12-\frac{5N}{2(N+1)}\Big] u_5-8\pi {\tilde u} u_5,
\eea
where $\tilde u$ is the dimensionless coupling constant of the quartic term defined in the main text and $l$ is the flow parameter. For $N>1/2$, $u_5$ is irrelevant at the GNY fixed point. Similar calculations also show that $u_6|\phi|^6$ and and $u^\prime_6 (\phi^6+h.c.)$ terms are irrelevant at the GNY fixed point. Thus, these terms can be safely neglected near the quantum phase transition for analyzing the FIQCP. \\

\begin{figure*}
\includegraphics[width=17.4cm]{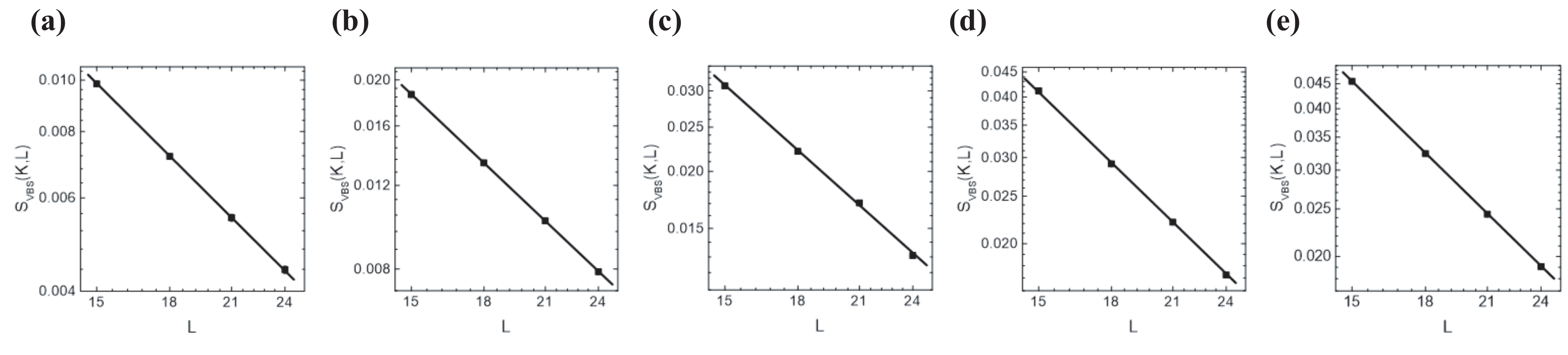}
\caption{ {$\mid$ \bf The critical exponent \boldsymbol{$\eta$}.} The critical exponent $\eta$ can be obtained from the fitting in the log-log plot of the Kekule-VBS structure factor versus $L$ at $J=J_c$ for different $N$.  ({\bf a}) $N=2$. ({\bf b}) $N=3$. ({\bf c}) $N=4$. ({\bf d}) $N=5$. ({\bf e}) $N=6$. \\  }
\label{figs2}
\end{figure*}

{\bf Supplementary Note 2: MQMC results of the \boldsymbol{$SU(N)$} models for \boldsymbol{$N=2,3,4,5,6$}}\\

In order to confirm the scenario of FIQCP, we perform sign-problem-free MQMC simulations of the $SU(N)$ fermionic model (Eq. 6) on honeycomb lattice for $N=2,3,4,5,6$. The largest linear system sizes in the simulations are $L=24$. We compute Binder ratio to determine phase transition points from semimetal to kekule-VBS phase. The results of Binder ratio for $N=2,4,5,6$ are shown in Supplementary Figure 1, from which we obtain the quantum critical values of $J$ for different $N$(shown in Fig. 2). We can see that the critical points of $J$ decrease monotonously as $N$ is increased. This trend is expected because the quantum fluctuations are stronger for smaller $N$ such that the critical values of $J$ are larger than the mean-field value. The critical $J_c$ in the limit of $N\to\infty$ is equal to the mean-field value.

\begin{figure*}
\includegraphics[width=17.4cm]{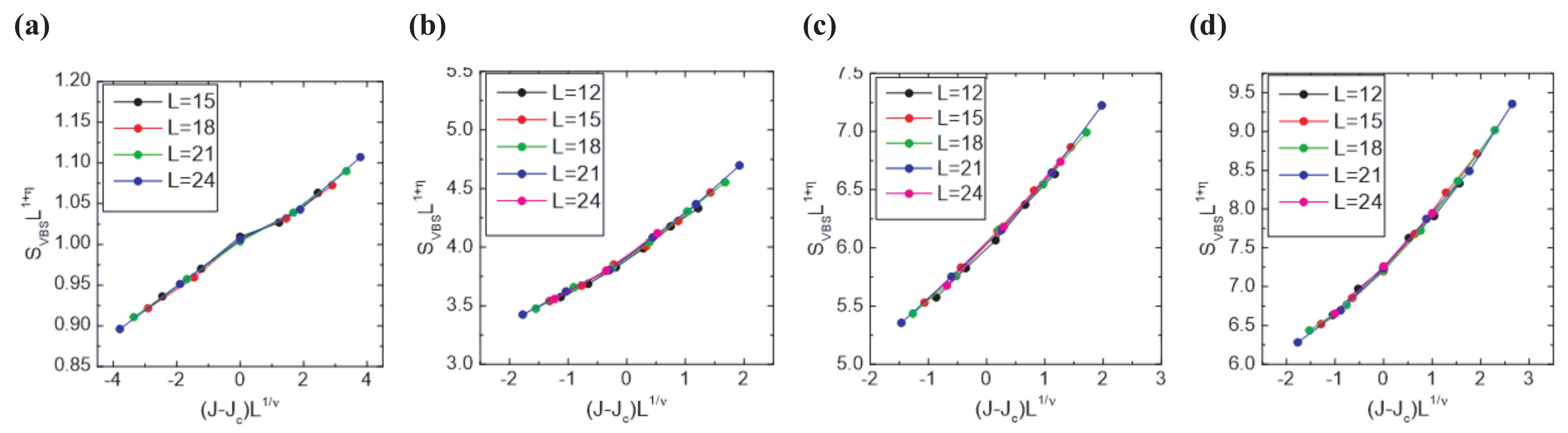}
\caption{{$\mid$ \bf  Data collapse results for different \boldsymbol{$N$}.} The critical exponent $\nu$ can be obtained through data collapse. ({\bf a}) $N=2$. ({\bf b}) $N=4$. ({\bf c}) $N=5$. ({\bf d}) $N=6$.   }
\label{figs3}
\end{figure*}

To study the critical behaviour of these transition points for different $N$, we use the scaling function $S_{\text{VBS}}({\bf K},L) = L^{-z-\eta} {\cal F}( L^{1/\nu}( J - J_c ))$ to obtain the critical exponents $\eta$ and $\nu$. The results are shown in Supplementary Figure 2 and Supplementary Figure 3. It is obvious that for $N=2,4,5,6$, the structure factor with different $L$ and $J$ can be collapsed to a smooth scaling function by choosing appropriate values of $\eta$ and $\nu$, which indicates that these transitions are continuous. The critical exponents $\eta$ and $\nu$ of the quantum phase transitions between the Dirac semimetals and the Kekule-VBS for $N=2,3,4,5,6$ fermions are summarized in Table. I of the main text.

In order to verify the prediction of emergent $U(1)$ symmetry at the FIQCP obtained in our RG's analysis, we employ the technique of histogram by studying the concurrence probability of VBS order parameter $P(\text{Re}(\phi),\text{Im}(\phi))$. The VBS order parameter $\phi$ is given by: $\phi = \sum_i (c^\dagger_i c_{i+\delta} + h.c) e^{i 2{\bf K} \cdot {\bf r}_i}$. Our histogram analysis shows that in the Kekule-VBS ordered phase, the concurrence probability $P(\text{Re}(\phi),\text{Im}(\phi))$ exhibits the expected $C_3$ symmetry. However, at the phase  transition point between the Dirac semimetal and the Kekule-VBS phase, the histogram of the VBS order parameter should exhibit an emergent $U(1)$ symmetry if the transition is a continuous one. The QMC results of the histogram both at the transition points and deep in the VBS ordered phase for $N=2,3$ are shown in Supplementary Figure 4, respectively. Indeed, the histogram at the transition point exhibit an emergent $U(1)$ symmetry, which provides strong evidences of ruling out the possibility of a first-order transition.

\begin{figure*}
\includegraphics[width=17.4cm]{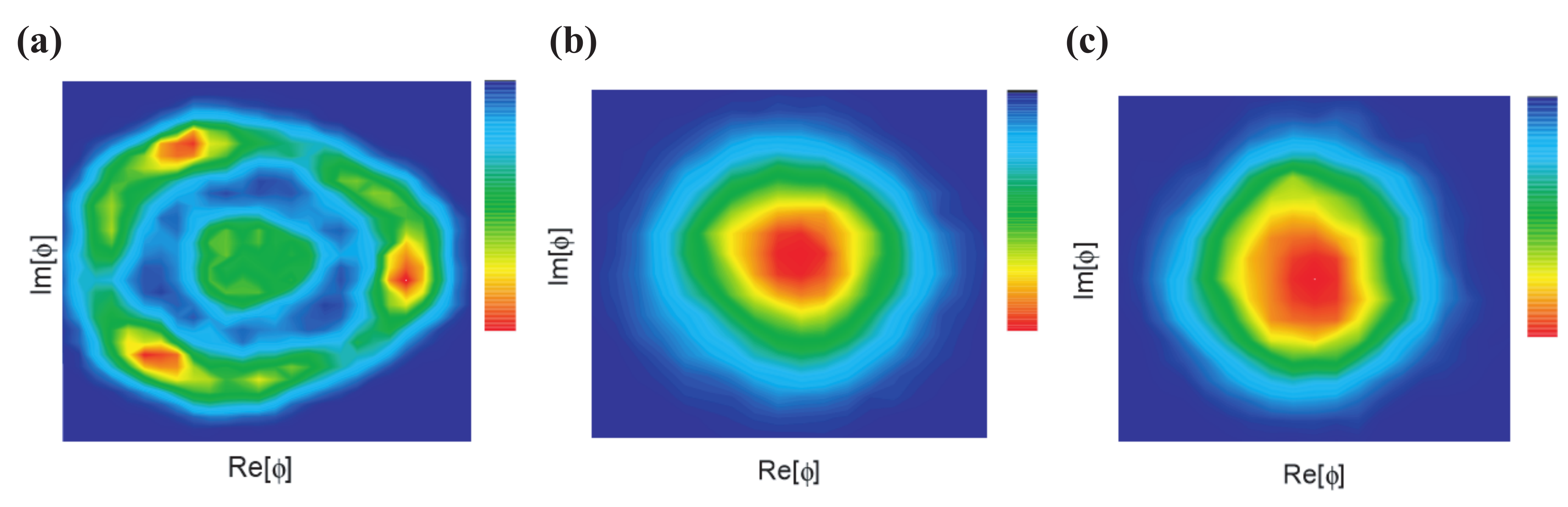}
\caption{{$\mid$ \bf The color-coded histogram of the VBS order parameter.} The color-coded histogram of the VBS order parameter in the system with $L=15$. Red color means the concurrence probability of VBS order parameter $P(\text{Re}(\phi),\text{Im}(\phi))$ is high, while blue color means concurrence probability of VBS order parameter $P(\text{Re}(\phi),\text{Im}(\phi))$ is low. ({\bf a}) $J/t=4.0$ for $N=3$: the system is located in the VBS ordered phase. The VBS order parameter exhibits $C_3$ symmetry. ({\bf b}) $J/t=2.95$ for $N=3$: the system is located at transition point from the Dirac semimetal to the Kekule-VBS phase. The VBS order parameter exhibits an emergent $U(1)$ symmetry. ({\bf c}) $J/t=3.2$ for $N=2$: the system is located at transition point from semimetal to VBS phase, at which an emergent $U(1)$ symmetry occurs.}
\label{figs4}
\end{figure*}

\end{widetext}


\begin{thebibliography}{99}

\bibitem{Subirbook} Sachdev, S. {\it Quantum Phase Transitions} (Cambridge U-
niversity Press, Cambridge, Ed. 2, 2011).
\bibitem{Landau-99} Landau, L. D., Lifshitz, E. M. \& Pitaevskii, E. M.  {\it Statistical Physics} (Butterworth-Heinemann, New York, 1999).
\bibitem{Wilson-74} Wilson, K. G.  \& Kogut, J. The renormalization group and the $\epsilon$ expansion, {\it Phys, Rep.} {\bf 12}, 75 (1974).
\bibitem{Sondhi-RMP} Sondhi, S. L., Girvin, S. M., Carini, J. P. \& Shahar, D.  Continuous quantum phase transitions, {\it Rev. Mod. Phys.} {\bf 69}, 315-333 (1997).
\bibitem{XGWenbook} Wen, X.-G.  {\it Quantum Field Theory of Many-body Systems}, (Oxford
    University Press, New York, 2004).
\bibitem{Fradkinbook} Fradkin, E.  {\it  Field Theories of Condensed Matter Physics}, (Cambridge University Press, Cambridge, Ed. 2, 2013).

\bibitem{Senthil-04} Senthil, T., Vishwanath, A., Balents, L., Sachdev, S. \& Fisher, M. P. A. Deconfined Quantum Critical Points.  {\it Science} {\bf 303}, 1490-1494 (2004).
\bibitem{Sandvik-2007} Sandvik, A. W.  Evidence for Deconfined Quantum Criticality in a Two-Dimensional Heisenberg Model with Four-Spin Interactions. {\it Phys. Rev. Lett.} {\bf 98}, 227202 (2007).
\bibitem{Melko-2008} Melko, R. G. \& Kaul, R. K. Scaling in the Fan of an Unconventional Quantum Critical Point.  {\it Phys. Rev. Lett.} {\bf 100}, 017203 (2008).
\bibitem{Dunghai-2010} Ghaemi, P., Ryu, S.  \& Lee, D.-H. Quantum valley Hall effect in proximity-induced superconducting graphene: An experimental window for deconfined quantum criticality.  {\it Phys. Rev. B} {\bf 81}, 081403(R) (2010).
\bibitem{Shao-16} Shao, H., Guo, W.  \& Sandvik, A. W. Quantum criticality with two length scales. {\it Science} {\bf 352}, 213-216 (2016).

\bibitem{Wu-82} Wu, F. Y. The Potts model.  {\it Rev. Mod. Phys.} {\bf 54}, 235-268 (1982).

\bibitem{Novoselov-05} Novoselov, K. S. A.,  Geim, A. K., Morozov, S. V., Jiang, D., Katsnelson, M. I., Grigorieva, I. V., Dubonos, S. V. \& Firsov, A. A. Two-dimensional gas of massless Dirac fermions in graphene.  {\it Nature} {\bf 438}, 197-200 (2005).

\bibitem{Neto-09} Castro Neto, A. H., Guinea, F., Peres, N. M. R., Novoselov, K. S.  \& Geim, A. K. The electronic properties of graphene.  {\it Rev. Mod. Phys.} {\bf 81}, 109-162 (2009).
\bibitem{Herbut-06} Herbut, I. F. Interactions and Phase Transitions on Graphene’s Honeycomb Lattice. {\it Phys. Rev. Lett.} {\bf 97}, 146401 (2006).
\bibitem{Assaad-13} Assaad, F. F.  \& Herbut, I. F. Pinning the Order: The Nature of Quantum Criticality in the Hubbard Model on Honeycomb Lattice.  {\it Phys. Rev. X} {\bf 3}, 031010 (2013).
\bibitem{Roy-13} Roy, B., Juricic, V. \& Herbut, I. F. Quantum superconducting criticality in graphene and topological insulators.  {\it Phys. Rev. B} {\bf 87}, 041401(R) (2013).
\bibitem{Vafek-14} Vafek, O. \& Vishwanath, A. Dirac Fermions in Solids: From High-Tc Cuprates and Graphene to Topological Insulators and Weyl Semimetals.  {\it Annual Review of Condensed Matter Physics} {\bf 5}, 83-112 (2014).

\bibitem{Moessner-01} Moessner, R., Sondhi, S. L. \& Chandra, P. Phase diagram of the hexagonal lattice quantum dimer model.  {\it Phys. Rev. B} {\bf64}, 144416 (2001).
\bibitem{Chamon-2007} Hou, C.-Y.,  Chamon, C. \& Mudry, C. Electron Fractionalization in Two-Dimensional Graphenelike Structures.  {\it Phys. Rev. Lett.} {\bf 98}, 186809 (2007).
\bibitem{Ryu-2009} Ryu, S., Mudry, C., Hou, C.-Y. \& Chamon, C. Masses in graphenelike two-dimensional electronic systems: Topological defects in order parameters and their fractional exchange statistics.   {\it Phys. Rev. B} {\bf 80}, 205319 (2009).
\bibitem{Pujari-13} Pujari, S., Damle, K. \& Alet, F. N\'eel-State to Valence-Bond-Solid Transition on the Honeycomb Lattice: Evidence for Deconfined Criticality. {\it Phys. Rev. Lett.} {\bf 111}, 087203 (2013).

\bibitem{Scalpino-81} Blankenbecler, B. \& Scalapino, D. J. \& Sugar, R. L. Monte Carlo calculations of coupled boson-fermion systems. I.  {\it Phys. Rev. D} {\bf 24}, 2278 (1981).
\bibitem{Fucito-81} Fucito, F., Marinari, E., Parisi, G. \& Rebbi, C. A proposal for Monte Carlo simulations of fermionic systems.  {\it Nucl. Phys. B} {\bf 180}, 369-377 (1981).
\bibitem{Hirsch-81} Hirsch, J. E., Scalapino, D. J., Sugar, R. L. \& Blankenbecler, R. Efficient Monte Carlo Procedure for Systems with Fermions.  {\it Phys. Rev. Lett.} {\bf47}, 1628 (1981).
\bibitem{Assaad-2005} Assaad, F. F. \& Evertz, H. G. {\it Computational Many-Particle Physics}, 277-356 (Lect. Notes Phys. 739, Springer, 2008).

\bibitem{ZXLi-15a} Li, Z.-X., Jiang, Y.-F. \& Yao, H. Solving the fermion sign problem in quantum Monte Carlo simulations by Majorana representation.  {\it Phys. Rev. B} {\bf 91}, 241117(R) (2015).

\bibitem{Rosenstein-93} Rosenstein, B., Yu, H. -L.   \&  Kovner, A.
Critical exponents of new universality classes.  {\it Phys. Lett. B.} {\bf 314}, 381-386 (1993). 

\bibitem{Moshe-03} Moshe, M. \& Zinn-Justin. J.
Quantum field theory in the large N limit: a review. {\it Phys. Rept.} {\bf385}, 69-228 (2003).

\bibitem{Pollmann-2015} Motruk, J., Grushin, A. G., de Juan, F \& Pollmann, F. Interaction-driven phases in the half-filled honeycomb lattice: An infinite density matrix renormalization group study.  {\it Phys. Rev. B} {\bf 92}, 085147 (2015).
\bibitem{Wu-2015} Zhou, Z., Wang, D., Meng, Z. Y., Wang, Y. \& Wu, C.  Mott insulating states and quantum phase transitions of correlated SU(2N) Dirac fermions.  {\it Phys. Rev. B} {\bf 93}, 245157 (2016).

\bibitem{Grover-14} Grover, T., Sheng, D. N. \& Vishwanath, A. Emergent Space-time Supersymmetry at the Boundary of a Topological Phase.  {\it Science} {\bf 344}, 280-283 (2014).
\bibitem{SSLee-14} Ponte, P. \& Lee, S.-S. Emergence of supersymmetry on the surface of three-dimensional topological insulators. {\it New J. of Physics}, {\bf 16}, 013044 (2014).
\bibitem{Jian-15} Jian, S.-K., Jiang, Y.-F. \& Yao, H. Emergent Spacetime Supersymmetry in 3D Weyl Semimetals and 2D Dirac Semimetals.  {\it Phys. Rev. Lett.} {\bf 114}, 237001 (2015).

\bibitem{Affleck-1988} Affleck, I. \& Marston, J. B.  Large-n limit of the Heisenberg-Hubbard model: Implications for high-${T}_{c}$ superconductors. {\it Phys. Rev. B} {\bf 37}, 3774 (1988).

\bibitem{Sachdev-1989} Read, N. \& Sachdev, S. Some features of the phase diagram of the square lattice SU(N) antiferromagnet. {\it Nucl. Phys. B} {\bf 316}, 609-640 (1989).

\bibitem{Wu-2003} Wu, C., Hu, J.-P.  \& Zhang, S.-C. Exact SO(5) Symmetry in the Spin-$3/2$ Fermionic System. {\it Phys. Rev. Lett.} {\bf 91}, 186402 (2003).

\bibitem{Honerkamp-2004}  Honerkamp, C. \& Hofstetter, W. Ultracold Fermions and the $\mathrm{SU}(N)$ Hubbard Model. {\it Phys. Rev. Lett.} {\bf 92}, 170403 (2004).

\bibitem{Troyer-2011} Corboz, P., L\"auchli, A. M., Penc, K., Troyer, M. \& Mila, F. Simultaneous Dimerization and SU(4) Symmetry Breaking of 4-Color Fermions on the Square Lattice.  {\it Phys. Rev. Lett.} {\bf 107}, 215301 (2011).

\bibitem{Congjun-2013} Cai, Z., Hung, H.-H., Wang, L., Zheng, D. \& Wu, C. Pomeranchuk Cooling of $\mathrm{SU}(2N)$ Ultracold Fermions in Optical Lattices. {\it Phys. Rev. Lett.} {\bf 110}, 220401 (2013).

\bibitem{Assaad-2013} Lang, T. C., Meng, Z. Y., Muramatsu, A., Wessel, S. \& Assaad, F. F. Dimerized Solids and Resonating Plaquette Order in $\mathrm{SU}(N)$-Dirac Fermions.  {\it Phys. Rev. Lett.} {\bf 111}, 066401 (2013).
    
\bibitem{Block-13} Block, M. S., Melko, R. G.  \& Kaul, R. K. Fate of $\mathbb{C}{\mathbb{P}}^{N\ensuremath{-}1}$ Fixed Points with $q$ Monopoles. {\it Phys. Rev. Lett.} {\bf 111}, 137202 (2013).

\bibitem{Loh-1990} Loh, E. Y., Gubernatis, J. E., Scalettar, R. T., White, S. R., Scalapino, D. J., \& R. L. Sugar, Sign problem in the numerical simulation of many-electron systems. {\it Phys. Rev. B} {\bf 41}, 9301 (1990).
\bibitem{Wu-2005} Wu, C. \& Zhang, S.-C. Sufficient condition for absence of the sign problem in the fermionic quantum Monte Carlo algorithm.  {\it Phys. Rev. B} {\bf 71}, 155115 (2005).
\bibitem{Troyer-2005} Troyer, M. \& Wiese, U.-J. Computational Complexity and Fundamental Limitations to Fermionic Quantum Monte Carlo Simulations. {\it Phys. Rev. Lett.} {\bf 94}, 170201 (2005).
\bibitem{Berg-2012} Berg, E., Metlitski, M. A. \& Sachdev, S.  Sign-Problem-Free Quantum Monte Carlo of the Onset of Antiferromag-netism in Metals.  {\it Science} {\bf 338}, 1606-1609 (2012).
\bibitem{ZXLi-2016sign} Li, Z.-X., Jiang Y.-F. \& Yao H. Majorana-Time-Reversal Symmetries: A Fundamental Principle for Sign-Problem-Free Quantum Monte Carlo Simulations. {\it Phys. Rev. Lett.} {\bf 117}, 267002 (2016).
\bibitem{TXiang-2016sign} Wei, Z. C., Wu C., Li Y., Zhang, S. \& Xiang T. Majorana Positivity and the Fermion Sign Problem of Quantum Monte Carlo Simulations. {\it Phys. Rev. Lett.} {\bf 116}, 250601 (2016).
\bibitem{Lei-2015} Wang, L., Liu, Y. H., Iazzi, M., Troyer, M. \& Harcos, G. Split Orthogonal Group: A Guiding Principle for Sign-Problem-Free Fermionic Simulations.  {\it Phys. Rev. Lett.} {\bf 115}, 250601 (2015).

\bibitem{Sorella-1989} Sorella, S., Baroni, S., Car, R. \& Parrinello, M. Non-Fermi-Liquid Exponents of the One-Dimensional Hubbard Model. {\it Europhys. Lett.} {\bf 8}, 663 (1989).
\bibitem{White-1989} White, S. R., Scalapino, D. J., Sugar, R. L., Loh,  E. Y., Gubernatis, J. E. \&  Scalettar, R. T. Numerical study of the two-dimensional Hubbard model. {\it Phys. Rev. B} {\bf 40}, 506 (1989).

\bibitem{Sorella2016} Otsuka, Y., Yunoki, S.  \& Sorella, S. Universal Quantum Criticality in the Metal-Insulator Transition of Two-Dimensional Interacting Dirac Electrons.  {\it Phys. Rev. X}  {\bf 6}, 011029 (2016).

\bibitem{columbia}  Gutierrez, C.  {\it et~al.} Imaging chiral symmetry breaking from Kekulé bond order in graphene. {\it Nature Physics} {\bf 12}, 950-958 (2016).

\bibitem{Moncton-1977} Moncton, D. E., Axe, J. D. \& DiSalvo, F. J. Neutron scattering study of the charge-density wave transitions in $2H\textrm{-}\mathrm{Ta}{\mathrm{Se}}_{2}$ and $2H\textrm{-}\mathrm{Nb}{\mathrm{Se}}_{2}$. {\it Phys. Rev. B} {\bf 16}, 801 (1977).

\bibitem{Herbut2016} Scherer, M. M. \& Herbut, I. F. Gauge-field-assisted Kekul\'e quantum criticality.  {\it Phys. Rev. B} {\bf 94}, 205136 (2016).
\bibitem{Jian2016} Jian, S.-K. \& Yao, H.  Fermion-induced quantum critical points in 3D Weyl semimetals. Preprint at http://arxiv.org/abs/1609.06313.

\bibitem{Cenke-2012} Xu, C. Unconventional Quantum Critical Point. {\it Int. J. Mod. Phys. B.} {\bf 26}, 1230007 (2012).

\end{thebibliography}
\end{document}